\documentclass[aps,prl,superscriptaddress,twocolumn]{revtex4-1}

\usepackage{amsmath}
\usepackage{color}
\usepackage{ulem}
\usepackage{hyperref}
\usepackage{graphicx,subfigure}

\def \be {\begin{equation}}
\def \ee {\end{equation}}
\def \ba {\begin{array}}
\def \ea {\end{array}}
\def \bea {\begin{eqnarray}}
\def \eea {\end{eqnarray}}

\def \ble {\begin{widetext}\begin{equation}}
\def \ele {\end{equation}\end{widetext}}
\def \blea {\begin{widetext}\begin{eqnarray}}
\def \elea {\end{eqnarray}\end{widetext}}

\def \nn {\nonumber}

\def \a {\alpha}
\def \b {\beta}
\def \g {\gamma}

\def \G {\Gamma}
\def \d {\delta}
\def \D {\Delta}
\def \e {\epsilon}

\def \n {\nu}

\def \l {\lambda}

\def \s {\sigma}

\def \r {\rho}

\def \O {\Omega}

\def \mA {\mathcal A}
\def \mB {\mathcal B}

\def \mD {\mathcal D}

\def \mO {\mathcal O}

\def \mT {\mathcal T}

\def \mX {\mathcal X}
\def \mY {\mathcal Y}

\def \cA {{\mathcal A}}

\def \cT {{\mathcal T}}

\def \cX {{\mathcal X}}
\def \cY {{\mathcal Y}}

\def \rL {{\mathrm L}}

\def \p {\partial}
\def \f {\frac}
\def \df {\dfrac}

\def \lt {\left}
\def \rt {\right}

\def \sr {\sqrt}

\def \inf {\infty}

\def \lag {\langle}
\def \rag {\rangle}

\def \ep {\mathrm{e}}
\def \ii {\mathrm{i}}

\def \tr {\mathrm{tr}}

\def \and {{\mathrm{and}}}

\def \rL {{\mathrm{L}}}
\def \NL {{\mathrm{NL}}}

\def \holo {{\rm{holo}}}

\def \CFT {{\rm{CFT}}}

\def \BO {{\rm{BO}}}

\def \Trho {\langle T \rangle_\rho}
\def \Arho {\langle \mathcal A \rangle_\rho}

\def \Trhoi {\langle T \rangle_{\rho_i}}
\def \Arhoi {\langle \mathcal A \rangle_{\rho_i}}
\def \Brhoi {\langle \mathcal B \rangle_{\rho_i}}
\def \Drhoi {\langle \mathcal D \rangle_{\rho_i}}

\def \Xrhoi {\langle \mathcal X \rangle_{\rho_i}}
\def \Yrhoi {\langle \mathcal Y \rangle_{\rho_i}}

\begin{document}

\title{Distinguishing Black Hole Microstates using Holevo Information}
%\title{Holevo Information of Thermal State in Two-dimensional Conformal Field Theory}

\author{Wu-zhong Guo}
\email{wzguo@cts.nthu.edu.tw}
\affiliation{Physics Division, National Center for Theoretical Sciences, National Tsing Hua University,\\No.\ 101, Sec.\ 2, Kuang Fu Road, Hsinchu 30013, Taiwan}

\author{Feng-Li Lin}
\email{linfengli@phy.ntnu.edu.tw}
\affiliation{Department of Physics, National Taiwan Normal University,\\No.\ 88, Sec.\ 4, Ting-Chou Road, Taipei 11677, Taiwan}

\author{Jiaju Zhang}
\email{jiaju.zhang@unimib.it}
\affiliation{Dipartimento di Fisica G.\ Occhialini, Universit\`a degli Studi di Milano-Bicocca,\\Piazza della Scienza 3, 20126 Milano, Italy}
\affiliation{INFN, Sezione di Milano-Bicocca, Piazza della Scienza 3, 20126 Milano, Italy}

\begin{abstract}
  We use the Holevo information in a two-dimensional conformal field theory (CFT) with a large central charge $c$ to distinguish microstates from the underlying thermal state. Holographically, the CFT microstates of a thermal state are dual to black hole microstate geometries in three-dimensional anti-de Sitter space. It was found recently that the holographic Holevo information shows plateau behaviors at both short and long interval regions. This indicates that the black hole microstates are indistinguishable from the thermal state by measuring over a small region, and perfectly distinguishable over a region with its size comparable to the whole system. In this letter, we demonstrate that the plateaus are lifted by including the $1/c$ corrections from both the vacuum and non-vacuum conformal families of CFT in either the canonical ensemble or microcanonical ensemble thermal state. Our results imply that the aforementioned indistinguishability and distinguishability of black hole microstate geometries from the underlying black hole are spoiled by higher order Newton constant $G_N$ corrections of quantum gravity.
\end{abstract}

\maketitle

%\tableofcontents

\section{Introduction}

The black hole information paradox lies in the fact that a pure state seems to evolve into a thermal state through Hawking radiation, and thus it violates unitarity of quantum mechanics. This paradox can be partially resolved if there exists black hole microstates, which are pure states, cannot be distinguished from the underlying thermal state.  This resolution however calls for a complete theory of quantum gravity which is beyond the reach at this moment. However, with the help of the anti-de Sitter/conformal field theory (AdS/CFT) correspondence \cite{Maldacena:1997re} one may glimpse the answer for this quantum gravity problem from the viewpoint of its dual CFT.

Recently, it was proposed in \cite{Bao:2017guc} to characterize distinguishability of the black hole microstates from its underlying thermal state by the Holevo information. One can call it in short the distinguishability of black hole microstates. The thermal state of the whole system is described by
\be \label{thermalstate}
\r = \sum_i p_i \r_i, ~~ \r_i = |i\rag \lag i|,
\ee
with the orthonormal microstates $|i\rag$ satisfying $\lag i|i'\rag = \d_{ii'}$. Note that $0 \leq p_i \leq 1$, $\sum_i p_i = 1$.
One would like to distinguish the microstates from the thermal state by performing measurements in a subsystem $A$, whose complement is denoted by $B$. The first step is to consider the relative entropy by comparing the reduced density matrix $\r_{A,i}=\tr_B \r_i$ of each of the microstates with the reduced density matrix $\r_{A}=\tr_B \r$ of the corresponding thermal state, i.e.,
\be
S(\r_{A,i}\|\r_A) = \tr(\r_{A,i}\log \r_{A,i}) - \tr(\r_{A,i}\log \r_{A}).
\ee
This quantity is a well-defined divergence and characterizes the difference between the two reduced density matrices.
The average relative entropy gives the Holevo information
\be
\chi_A = \sum_i p_i S(\r_{A,i}\|\r_A) = S_A - \sum_i p_i S_{A,i},
\ee
with entanglement entropies (EEs) $S_A = -\tr (\r_A \log \r_A)$, $S_{A,i} = -\tr (\r_{A,i} \log \r_{A,i})$.
It is just the difference between the thermal state EE and the average EE of the microstates.
The Holevo information $\chi_A$ is the upper bound of the mutual information between the thermal state and any measurement inside $A$, which is aiming to reproduce the thermal state and to characterize the accessible information.

By construction
\be
0 \leq \chi_A \leq S_{\rm thermal},
\ee
with $S_{\rm thermal}$ being thermal entropy of the whole system
\be
S_{\rm thermal} = - \sum_i p_i \log p_i.
\ee
When $\chi_A=0$, $\r_{A,i} = \r_{A}$ so that the microstates are totally indistinguishable by measurements inside $A$. On the other hand, when $\chi_A=S_{\rm thermal}$, $\r_{A,i} \r_{A,i'} = 0$ for arbitrary $i,i'$ and thus the microstates are completely distinguishable.

To investigate the information loss paradox of black hole in Einstein gravity in the AdS$_3$ background, i.e., the Ba\~nados-Teitelboim-Zanelli (BTZ) black hole \cite{Banados:1992wn}, we calculate the Holevo information in a two-dimensional (2D) CFT. When the gravity is weakly coupled, the CFT has a large central charge \cite{Brown:1986nw}
\be c=\f{3 R}{2 G_N}, \ee
with $G_N$ being the Newton constant and $R$ being the AdS radius. The $1/c$ corrections on the CFT side correspond to quantum corrections on the gravity side.

We consider a 2D large $c$ CFT in thermal state on a cylinder with spatial period $L$. For an interval $A$ with length $\ell$, we denote the Holevo information by $\chi(\ell)$.
The Holevo information $\chi(\ell)$ is monotonically increasing with respect to $\ell$. It is easy to see that
\be
\lim_{\ell\to0}\chi(\ell)=0, ~~ \lim_{\ell\to L}\chi(\ell)=S(L).
\ee

By using the holographic entanglement entropy (HEE) \cite{Ryu:2006bv,Hubeny:2007xt}, it was recently found in \cite{Bao:2017guc} that the holographic Holevo information shows plateau behaviors around both $\ell\to0$ and $\ell\to L$. This indicates that the microstates are totally indistinguishable until the interval reaches a non-vanishing critical length, and are perfectly distinguishable after the interval reaches another critical length that is shorter than length of the whole system.
However, the HEE is only the classical gravity result, and it is expected that quantum corrections to the HEE \cite{Headrick:2010zt,Barrella:2013wja,Faulkner:2013ana} would resolve both plateaus of the holographic Holevo information. On the dual CFT side, these correspond to $1/c$ corrections.
The problem has been addressed in \cite{Michel:2018yta} for the 2D CFT due to the zero mass BTZ black hole.
In this letter, we consider the more general thermal states, including the canonical ensemble thermal state with both high and low temperatures, as well as the microcanonical ensemble thermal state. This is not only technically challenging by performing the thermal average over all eigenstates, i.e., including both primaries and their descendants, but also conceptually interesting to see if the peculiar non-thermal/non-geometrical descendants states found in \cite{Guo:2018fnv} will be thermally averaged out so that the microstates remain almost ultra-locally indistinguishable.

We find that the Holevo information is not vanishing as long as the length of the interval is non-vanishing, and this indicates that the black hole microstates are distinguishable from thermal state as long as the measuring region is non-vanishing.  We also find the Holevo information is smaller than the thermal entropy as long as the interval is shorter than the whole system.

For calculation convenience we choose that the interval $A$ is short, i.e.,  $\ell / L \ll 1$, and thus its complement $B$ has a length $L-\ell$ comparable to $L$. Then we have
\bea
&& S_A = S(\ell), ~~
   S_{A,i} = S_i(\ell), ~~
   \chi_A = \chi(\ell),\\
&& S_B = S(L-\ell), ~~
   S_{B,i} = S_i(L-\ell), ~~
   \chi_B = \chi(L-\ell).\nn
\eea
Note that $S_{A,i}=S_{B,i}$. To get the short and long interval Holevo information $\chi_A$ and $\chi_B$, we need to calculate the short and long interval EEs of thermal state, i.e., $S_A$, $S_B$, and the average of the short interval EEs of the microstates, i.e., $\sum_i p_i S_{A,i}$.
For the short interval, as in \cite{Chen:2016lbu,Lin:2016dxa,He:2017vyf,He:2017txy}, we use the operator product expansion (OPE) of twist operators \cite{Calabrese:2004eu,Cardy:2007mb,Headrick:2010zt,Calabrese:2010he,Rajabpour:2011pt,Chen:2013kpa,Bianchini:2015uea} to calculate the short interval expansion of the EE. This method is still available for the long interval case \cite{Chen:2014ehg,Chen:2015kua,Chen:2017ahf}.

\section{Canonical ensemble thermal state with high temperature}

\begin{figure*}[htpb]
  \centering
  \includegraphics[height=0.22\textwidth]{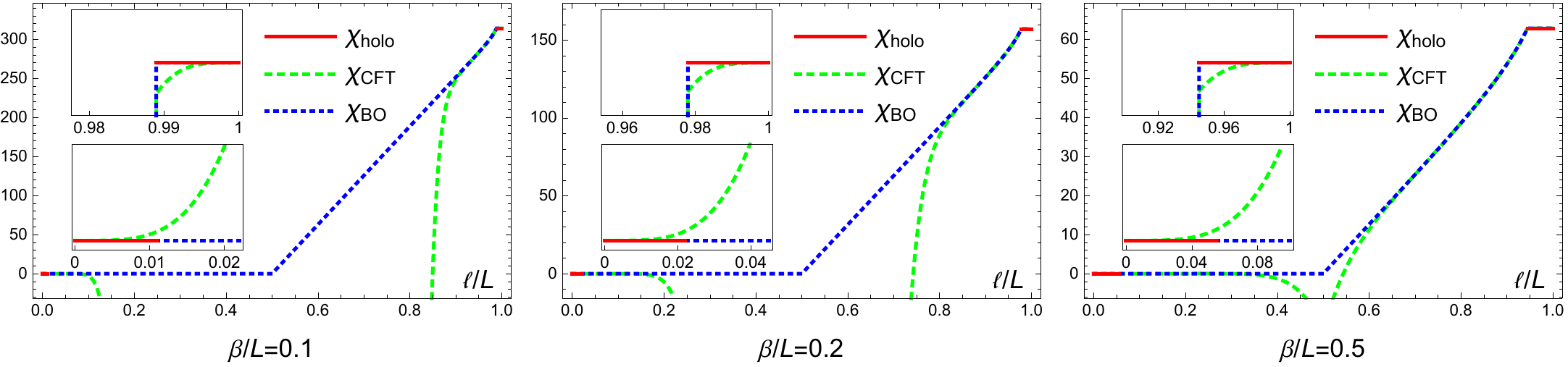}
  \caption{The holographic Holevo information $\chi_\holo$ (\ref{chiholo}), the short and long interval expansion of the CFT Holevo information $\chi_\CFT$ (\ref{chiA}) and (\ref{chiB}), i.e., (S14) and (S18) in the supplemental material, and the leading order $c$ Holevo information $\chi_\BO$ (\ref{chiBO}), for the high temperature thermal state with $\beta/L=0.1$ (Left), $\beta/L=0.2$ (Middle), and $\beta/L=0.5$ (Right), respectively. The unknown region of holographic Holevo information $\chi_\holo$ is left blank. To draw the figures we have set $c=30$.}\label{holoCFT}
\end{figure*}

For a canonical ensemble thermal state we have
\be
p_i = \f{\ep^{-\b E_i}}{Z(\b)}, ~~ Z(\b) = \sum_i \ep^{-\b E_i},
\ee
with $\b$ being the inverse temperature. We consider high temperature limit $\b/L\ll1$ and omit the terms suppressed by the exponential factor $\ep^{-2\pi L/\b}$. The thermal entropy is
\be
S(L) = \f{\pi c L}{3\b},
\ee
which is just the entropy of a non-rotating BTZ black hole.
Using the HEE \cite{Ryu:2006bv,Hubeny:2007xt}, one can get the holographic Holevo information \cite{Bao:2017guc}
\be \label{chiholo}
\chi_\holo(\ell) = \lt\{
\ba{cl}
0 & \ell < \f{\b}{2\pi}\log2 \\
\f{\pi c L}{3\b} & \ell> L - \f{\b}{2\pi}\log2
\ea
\rt.\!\!\!.
\ee
The holographic Holevo information $\chi_\holo(\ell)$ with $\f{\b}{2\pi}\log2 < \ell < L-\f{\b}{2\pi}\log2$ is unknown. The result is plotted in Fig.~\ref{holoCFT}. There are plateaus at both $\ell<\f{\b}{2\pi}\log2$ and $\ell>L-\f{\b}{2\pi}\log2$. We will resolve the plateaus in CFT.

We consider only contributions from the vacuum conformal family, and will briefly discuss  the contributions from non-vacuum conformal families in the end of the letter. For the short interval $A$ we have the EE \cite{Calabrese:2004eu}
\be \label{SAexact}
S_A = \f{c}{3}\log \Big( \f{\b}{\pi\e}\sinh\f{\pi\ell}{\b} \Big).
\ee
Though we do not calculate $S_{A,i}$ for all the pure states, using the results in \cite{He:2017txy,Guo:2018pvi} we can get the average EE
\bea \label{SipiSi}
&& \sum_i p_i S_{A,i} = \f{c}{3}\log\f{\ell}{\e}
  +\frac{\pi^2 c \ell^2}{18 \beta^2}
  -\frac{\pi^3\ell^4 (\pi  c L +24 \beta)}{540 \beta^4 L}\\
&& ~~ ~~ ~~
  +\frac{\pi^4 \ell^6(\pi^2 c^2 L^2+72 \pi c \beta L+864 \beta^2)}{8505 c \beta^6 L^2}
  + \cdots
  + O(\ell^{12}).\nn
\eea
We have omitted some involved terms denoted by $\cdots$, and one can find full form of the equation in (S13) of the supplemental material.
There are technical issues in calculating the result to higher orders of $\ell$. See details in the supplemental material.
Combining them, we obtain the short interval Holevo information
\be\label{chiA}
\chi_A = \frac{2 \pi^3 \ell^4}{45 \beta^3 L}
       - \frac{8\pi^4\ell^6 (\pi  c L + 12 \beta ) }{945 c \beta^5 L^2}
       + \cdots
       + O(\ell^{12}).
\ee
See full form of the equation in (S14) of the supplemental material.
We find that to the order we consider it is vanishing in the thermodynamic limit \cite{Lashkari:2016vgj,Dymarsky:2016aqv}, i.e., the limit $L \rightarrow \infty$ with $\b,\ell$ fixed.

For the long interval $B$ we have the EE \cite{Chen:2017ahf}
\be
S_B =  \f{c}{3} \log \Big( \f{\b}{\pi\e} \sinh\f{\pi\ell}{\b} \Big)
     + \f{\pi c L}{3\b}
     - I(1-\ep^{-\f{2\pi\ell}{\b}}).
\ee
The function $I(x)$ is the mutual information of two intervals on a complex plane with cross ratio $x$.
The small $x$ expansion of $I(x)$ to order $x^{8}$ was calculated in \cite{Barrella:2013wja,Chen:2013dxa} and to order $x^{10}$ was calculated in \cite{Beccaria:2014lqa,Li:2016pwu}.
Note that, nothing but tediousness prevents one from calculating the mutual information to even higher orders of $\ell$.
Combining with the fact $S_{B,i}=S_{A,i}$, we obtain the long interval Holevo information
\bea \label{chiB}
&& \chi_B = \frac{\pi  c L}{3 \beta }
          -\frac{2 \pi^3 ( 4 \pi  L - 7 \beta )\ell^4}{315 \beta^4 L}
          +\frac{32 \pi^5 \ell^5}{3465 \beta^5}  \\
&& \phantom{\chi_B =}
          +\frac{8 \pi^4 ( 32 \pi^2 L^2-143 \pi \beta L )\ell^6}{135135 \beta^6L^2}
          + \cdots
          +O(\ell^{11},1/c). \nn
\eea
One can find full form of the equation in (S18) of the supplemental material.
Note that $S(L)-\chi_B$ is non-vanishing in the thermodynamic limit.

We denote the results (\ref{chiA}) and (\ref{chiB}) as the CFT Holevo information $\chi_\CFT(\ell)$ and  $\chi_\CFT(L-\ell)$, respectively. Note that they are only valid for $\ell\ll\b\ll L$. They are consistent with the holographic Holevo information $\chi_\holo$ (\ref{chiholo}) at the leading order of large $c$, while at the sub-leading orders we see the corrections. We plot them in Fig.~\ref{holoCFT}. We see that with $1/c$ corrections both the short and long interval plateaus are resolved.

The leading $c$ of (\ref{SipiSi}) is consistent with the result
\be \label{SipiSic}
\sum_i p_i S_{A,i} = \f{c}{3}\log\Big( \f{\b}{\pi\e}\sinh\f{\pi\ell}{\b} \Big) + O(c^0),
\ee
which was got in \cite{Bao:2017guc} by assuming that the contributions from the primary excited states dominate the average.
In fact, from the result in \cite{Kraus:2016nwo}, we can show that there are far more descendant states than primary states in high levels of a large $c$ CFT \cite{Guo:2018pvi}. It is intriguing to show explicitly why primary excited states dominate the average.
Supposing (\ref{SipiSic}) is valid as long as $\ell<L/2$, one can get the Holevo information by Bao and Ooguri in \cite{Bao:2017guc}
\be \label{chiBO}
\chi_\BO(\ell) = \lt\{
\ba{cl}
0                                                                & \ell<L/2 \\
\f{c}{3} \log \f{\sinh\f{\pi\ell}{\b}}{\sinh\f{\pi(L-\ell)}{\b}} & L/2<\ell<L-\f{\b}{2\pi}\log2\\
\f{\pi c L}{3\b}                                                   & \ell > L-\f{\b}{2\pi}\log2
\ea
\rt.\!\!\!.
\ee
It is a combination of the holographic and CFT results, and is the leading order $c$ Holevo information.
For comparison, we also plot $\chi_\BO$ in Fig.~\ref{holoCFT}.

\section{Canonical ensemble thermal state with low temperature}

In low temperature limit, we have $\b\gg L$. The dual gravity background is the thermal AdS and the holographic thermal entropy is vanishing
\be
S_\holo(L)=0.
\ee
From $0\leq\chi(\ell)\leq S(L)$, we obtain
\be
\chi_\holo(\ell)=0.
\ee

In CFT, the above total indistinguishability can be lifted by taking into account the finite-size effect exponentially suppressed by the factor $q=\ep^{-2\pi\b/L}$. Using the results in \cite{Chen:2017ahf} and considering only the contributions from the holomorphic sector of the vacuum conformal family, for the short interval we get
\bea \label{chiAlow}
&& \chi_A = \Big[ \frac{32 q^2}{15 c}
                 +\frac{24 q^3}{5 c}
                 +\frac{64 q^4}{5 c}+O(q^5) \Big] \Big( \f{\pi\ell}{L} \Big)^4 \nn\\
&& \phantom{\chi_A =}
           +\Big[ \frac{128 (c-16) q^2}{315 c^2}
                 +\frac{32 (c-24) q^3}{35 c^2} \\
&& \phantom{\chi_A =}
                 +\frac{256 (c-40) q^4}{105 c^2}+O(q^5) \Big] \Big( \f{\pi\ell}{L} \Big)^6
           +O(\ell^8), \nn
\eea
and for the long interval we obtain
\bea \label{chiBlow}
&& \chi_B - S(L) = - \Big[ \f{32\pi\b(\b^2+L^2)(4\b^2+L^2)}{15L^5}q^2 \nn\\
&& \phantom{\chi_B  - S(L) =} + O(q^3) \Big] \Big( \f{\pi\ell}{L} \Big)^4 + O(\ell^5).
\eea

\section{Microcanonical ensemble thermal state}

We now consider the microcanonical ensemble thermal state with fixed high energy $E$, with contributions from both the holomorphic and  anti-holomorphic sectors. We have the thermal sate (\ref{thermalstate}) with
\be \label{pime}
p_i = \f{\d(E-E_i)}{\O(E)}.
\ee
At energy $E$ the number of states $\O(E)$ is given by the Cardy formula \cite{Cardy:1986ie} and it is an inverse Laplace transformation of canonical ensemble partition function $Z(\b)$.
Beyond the saddle point approximation of \cite{Cardy:1986ie,Carlip:2000nv}, it turns out that
\be \label{OE}
\O(E) = \sqrt{\f{\pi c L}{6E}} I_1\Big(\sqrt{\f{2\pi c L E}{3}}\Big),
\ee
with $I_\n$ being modified Bessel function of the first kind.
As the case of canonical ensemble thermal state with high temperature, we omit the exponentially suppressed terms of large $E$ but keep the power suppressed terms.

The Cardy formula can be generalized to the cases of various multi-point correlation functions on a torus \cite{Kraus:2016nwo,Brehm:2018ipf,Romero-Bermudez:2018dim,Hikida:2018khg}, i.e., in canonical ensemble thermal state. One can use the inverse Laplace transformation of the canonical ensemble average to obtain the corresponding microcanonical ensemble one.
In this way, we can derive the one-point functions, and thus the short interval EE, of the microcanonical ensemble thermal state from the canonical ensemble one-point functions. Similarly, we can obtain the microcanonical ensemble average short interval EE from the corresponding canonical ensemble one. Combining the short interval EE and average EE, we obtain the Holevo information
\be \label{chiAmc}
\chi_A = \frac{\pi ^3 \ell ^4[ \pi  c L ({I_3}-{I_1})+24 \lambda{I_2} ]}{540\lambda ^4 L{I_1}}
        + \cdots
        + O(\ell^{12}),
\ee
with the definition $\l := \sqrt{\f{\pi c L}{6E}}$, which is fixed in the thermodynamic limit, and $I_\n$ being the shorthand notation of $I_\n(\f{\pi c L}{3\l})$. The full form of the equation is presented in (S38) of the supplemental material.

For the long interval case, we use the OPE of twist operators in \cite{Chen:2014ehg,Chen:2015kua,Chen:2017ahf} and obtain the following result,
\be \label{chiBmc}
\chi_B - S(L) = O(\ell^{12}).
\ee
However, we cannot get the term of order $\ell^{12}$ explicitly. It is possibly non-vanishing. See details in the supplemental material.

\section{Contributions from a non-identity primary operator}

Lastly, we consider the leading contribution to the Holevo information from a non-identity primary operator $\psi$ with normalization $\a_\psi$, conformal weights $(h_\psi,\bar h_\psi)$. We have the scaling dimension $\D_\psi=h_\psi+\bar h_\psi$ and spin $s_\psi=h_\psi-\bar h_\psi$. For a general thermal state with density matrix (\ref{thermalstate}), we use the OPE of twist operators \cite{Calabrese:2004eu,Cardy:2007mb,Headrick:2010zt,Calabrese:2010he,Rajabpour:2011pt,Chen:2013kpa,%
Chen:2014ehg,Bianchini:2015uea,Chen:2015kua,Chen:2017ahf} and get the short and long interval Holevo information
\blea \label{dpsichiAdpsichiB}
&& \d_\psi \chi_A = \f{\sr\pi\G(\D_\psi+1)\ell^{2\D_\psi}}{2^{2\D_\psi+2}\G(\D_\psi+\f32)}
                 \f{\ii^{2s_\psi}}{\a_\psi}
                 \Big[ \sum_i p_i \lag \psi \rag_{\r_i}^2
                     - \Big( \sum_i p_i \lag \psi \rag_{\r_i} \Big)^2
                 \Big]
                 + o(\ell^{2\D_\psi}), \nn\\
&& \d_\psi \chi_B = \d_\psi S(L)
               - \f{\ell^{2\D_\psi}}{2^{2\D_\psi+1}}
                 \f{\ii^{2s_\psi}}{\a_\psi}
                 \sum_{i \neq i'} \lag i | \psi | i' \rag \lag i' | \psi | i \rag
                                  p_i \p_n \Big[ \sum_{j=1}^{n-1} \f{(p_{i'}/p_i)^j}{(\sin \f{\pi j}{n})^{2\D_\psi}} \Big]_{n=1}
                 + o(\ell^{2\D_\psi}).
\elea
These forms are general and can be applied to both canonical ensemble and microcanonical ensemble thermal states. The results however are not universal in the sense that they depend on the structure constants, so that we cannot evaluate their explicit forms without knowing the details of the theory.  See more details in the supplemental material.

\section{Discussion}

For concluding the letter, we would like to mention the implication of the almost vanishing short interval Holevo information to our recent finding of non-geometric states in \cite{Guo:2018fnv}. As shown in \cite{Guo:2018fnv} some special descendant states are non-geometric, which indicates that they cannot be locally like thermal. The ensemble average for obtaining the Holevo information is over all states including those non-geometric descendant states. However, we see the resultant leading order $c$ short interval Holevo information is still consistent with thermality. Using the results in \cite{Kraus:2016nwo} we can show there are far more descendant states than primary ones at high levels in a large $c$ CFT \cite{{Guo:2018pvi}}. This indicates that the contributions from the non-geometric descendant states are suppressed. It is intriguing to show how this happens explicitly.

~

\noindent\textit{%
We would like thank Alice Bernamonti, Pasquale Calabrese, Federico Galli, Manuela Kulaxizi, Hong Liu, Andrei Parnachev, Tadashi Takayanagi, and Erik Tonni for helpful discussions. JZ would like to thank the Galileo Galilei Institute for Theoretical Physics and the organisers of the workshop ``Entanglement in Quantum Systems'' for hospitality and for being given the opportunity to present part of the result of work, and to thank participants of the workshop for helpful discussions. WZG is supported in part by the National Center of Theoretical Science (NCTS). FLL is supported by Taiwan Ministry of Science and Technology through Grant No.~103-2112-M-003-001-MY3. JZ is supported in part by Italian Ministero dell'Istruzione, Universit\`a e Ricerca (MIUR), and Istituto Nazionale di Fisica Nucleare (INFN) through the ``Gauge Theories, Strings, Supergravity'' (GSS) research, and by Fondazione Cariplo and Regione Lombardia, Grant No.\ 2015-1253.%
}


%\bibliographystyle{D:/00.zbib/apsrev4-1}   %%utphys %%utcaps %%JHEP %%apsrev4-1
%\bibliography{D:/00.zbib/2018,D:/00.zbib/2017,D:/00.zbib/1970,D:/00.zbib/1980,D:/00.zbib/1990,D:/00.zbib/1995,D:/00.zbib/1996,D:/00.zbib/1997,D:/00.zbib/1998,D:/00.zbib/1999,D:/00.zbib/2000,D:/00.zbib/2001,D:/00.zbib/2002,D:/00.zbib/2003,D:/00.zbib/2004,D:/00.zbib/2005,D:/00.zbib/2006,D:/00.zbib/2007,D:/00.zbib/2008,D:/00.zbib/2009,D:/00.zbib/2010,D:/00.zbib/2011,D:/00.zbib/2012,D:/00.zbib/2013,D:/00.zbib/2014,D:/00.zbib/2015,D:/00.zbib/2016,D:/00.zbib/book,D:/00.zbib/work,D:/00.zbib/thesis}

%%%%%%%%%%%%%%%%%%%%%%%%%%%%%%%%%%%%%%%%%%%%%%%%%%%%%%%%%%%%%%%%%%
%% Begin supplemental material
%%%%%%%%%%%%%%%%%%%%%%%%%%%%%%%%%%%%%%%%%%%%%%%%%%%%%%%%%%%%%%%%%%

\newpage

%change the format of equations%

\renewcommand\thepage{{S}{\arabic{page}}}
\renewcommand\theequation{{S}{\arabic{equation}}}
\renewcommand\thefigure{{S}{\arabic{figure}}}

\setcounter{page}{1}
\setcounter{equation}{0}
\setcounter{figure}{0}

%\title{Distinguishing Black Hole Microstates using Holevo Information\\Supplemental Material}

%\author{Wu-zhong Guo}
%\email{wzguo@cts.nthu.edu.tw}
%\affiliation{Physics Division, National Center for Theoretical Sciences, National Tsing Hua University, %\\No.\ 101, Sec.\ 2, Kuang Fu Road, Hsinchu 30013,
%Hsinchu 30013, Taiwan}

%\author{Feng-Li Lin}
%\email{linfengli@phy.ntnu.edu.tw}
%\affiliation{Department of Physics, National Taiwan Normal University, %\\No.\ 88, Sec.\ 4, Ting-Chou Road, Taipei 11677,
%Taipei 11677, Taiwan}

%\author{Jiaju Zhang}
%\email{jiaju.zhang@unimib.it}
%\affiliation{Dipartimento di Fisica G.\ Occhialini, Universit\`a degli Studi di Milano-Bicocca, %\\Piazza della Scienza 3,
%20126 Milano, Italy}
%\affiliation{INFN, Sezione di Milano-Bicocca, %Piazza della Scienza 3, 20126
%20126 Milano, Italy}

%\begin{abstract}
%  This is the supplemental material to the letter \cite{Guo:2018djz}. We give the calculation details and the full forms of the omitted equations.
%\end{abstract}

%\maketitle

\begin{center}
{\large \textbf{Distinguishing Black Hole Microstates using Holevo Information\\Supplemental Material}}
\\~\\
{Wu-zhong Guo, Feng-Li Lin, and Jiaju Zhang}
\end{center}

%\tableofcontents

%\normalsize%\large%\Large

\begin{center}
{\normalsize \textbf{Canonical ensemble thermal state with high temperature}}
\end{center}

With contributions from only the vacuum conformal family, the EE of one short interval in a general state $\r$ can be written as \cite{He:2017txy,Guo:2018pvi}
\bea \label{SAOPE}
&& S_{A} = \f{c}{3}\log\f\ell\e + 2 \big[ \ell^2 a_T \Trho
                                  + \ell^4 a_{TT} \Trho^2
                                  + \ell^6 a_{TTT} \Trho^3 \nn\\
&& \phantom{S_{A}=}
           + \ell^8 \big( a_{\mA\mA} \Arho^2
                        + a_{TT\mA} \Trho^2\Arho
                        + a_{TTTT} \Trho^4 \big) \nn\\
&& \phantom{S_{A}=}
           + \ell^{10} \big( a_{T\mA\mA} \Trho\Arho^2
                           + a_{TTT\mA} \Trho^3\Arho \nn\\
&& \phantom{S_{A}=}
                           + a_{TTTTT} \Trho^5 \big)
                           +O(\ell^{12}) \big].
\eea
with the coefficients
\bea
&& a_T = -\f16, ~~
   a_{TT} = -\frac{1}{30 c}, ~~
   a_{TTT} = -\frac{4}{315 c^2}, \nn\\
&& a_{\mA\mA} = -\frac{1}{126 c (5 c+22)}, ~~
   a_{TT\mA} = \frac{1}{315 c^2}, \nn\\
&& a_{TTTT} = -\frac{c+8}{630 c^3}, ~~
   a_{T\mA\mA} = -\frac{16}{693 c^2 (5 c+22)}, \nn\\
&& a_{TTT\mA} = \frac{32}{3465 c^3}, ~~
   a_{TTTTT} = -\frac{16 (c+5)}{3465 c^4}.
\eea
Here $T$ is the stress tensor, and $\mA = (TT) - \f{3}{10}\p^2 T$. The density matrix $\r$ can be either a thermal state, or any individual pure state, and in fact it can be any state that is translational invariant. We have included the contributions from both the holomorphic and anti-holomorphic sectors, and it is applied to the states in which the contributions from the holomorphic and anti-holomorphic sectors are the same. Otherwise, we can just write the holomorphic and anti-holomorphic contributions separately.

In high temperature limit we omit the exponentially suppressed terms and get
\be \label{TbAb}
\lag T \rag_\b=-\f{\pi^2c}{6\b^2}, ~~
\lag \mA \rag_\b=\f{\pi^4c(5c+22)}{180\b^4},
\ee
which are just one-point functions on a cylinder with infinite space and temporal period $\b$.
Using (\ref{SAOPE}) and (\ref{TbAb}) we get the EE
\bea
&& S_A = \f{c}{3}\log\f{\ell}{\e}
    + \frac{\pi^2 c \ell^2}{18 \beta ^2}
    -\frac{\pi^4 c \ell^4}{540 \beta ^4}
    +\frac{\pi^6 c \ell^6}{8505 \beta^6}
    -\frac{\pi^8 c \ell^8}{113400 \beta ^8} \nn\\
&& \phantom{S_A =}
    +\frac{\pi^{10} c \ell^{10}}{1403325 \beta^{10}}
    +O(\ell^{12}),
\eea
which is consistent with the exact result \cite{Calabrese:2004eu}
\be
S_A = \f{c}{3}\log \Big( \f{\b}{\pi\e}\sinh\f{\pi\ell}{\b} \Big).
\ee

To calculate the average EE, we first calculate the average products of one-point functions
\begin{widetext}
\bea \label{SipiXi}
&& \sum_i p_i \Trhoi = -\frac{\pi^2 c}{6 \beta^2}, ~~
   \sum_i p_i \Trhoi^2 = \frac{\pi^3 c (\pi  c L + 24 \beta)}{36 \beta^4 L}, ~~
   \sum_i p_i \Trhoi^3 = -\frac{\pi^4 c (\pi^2 c^2 L^2+72 \pi c \beta L+864 \beta^2)}{216 \beta^6 L^2}, \nn\\
&& \sum_i p_i \Trhoi^4 = \frac{\pi^5 c (\pi^3 c^3 L^3+144 \pi^2 c^2 \beta  L^2+5184 \pi c \beta^2 L+41472 \beta^3)}
                              {1296 \beta^8 L^3}, \nn\\
&& \sum_i p_i \Trhoi^5 = -\frac{\pi^6 c (\pi^4 c^4 L^4+240 \pi^3 c^3 \beta L^3
                                        +17280 \pi^2 c^2 \beta^2 L^2+414720 \pi c \beta^3 L
                                        +2488320 \beta^4)}{7776 \beta^{10} L^4}, \nn\\
&& \sum_i p_i \Arhoi = \frac{\pi^4 c (5 c+22)}{180 \beta^4}, ~~
   \sum_i p_i \Trhoi\Arhoi = -\frac{\pi^5 c (5 c+22) (\pi  c L + 48 \beta)}{1080 \beta^6 L}, \nn\\
&& \sum_i p_i \Trhoi^2\Arhoi = \frac{\pi^6 c (5 c+22) (\pi^2 c^2 L^2+120 \pi c \beta L+2880 \beta^2)}{6480 \beta^8 L^2} ,\nn\\
&& \sum_i p_i \Trhoi^3\Arhoi = -\frac{\pi^7 c (5 c+22) (\pi^3 c^3 L^3+216 \pi^2 c^2 \beta  L^2
                                                       +12960 \pi c \beta^2L+207360 \beta^3)}{38880 \beta^{10} L^3}, \nn\\
&& \sum_i p_i \Arhoi^2 = \frac{\pi^7 c (5 c+22) [7 \pi  c (5 c+22) L +480 (7 c+74)\beta]}{226800 \beta^8 L}, \\
&& \sum_i p_i \Trhoi\Arhoi^2 = -\frac{\pi^8 c (5 c+22) [7 \pi^2 c^2 (5 c+22) L^2+192 \pi c  (35 c+262) \beta L
                                                       +40320 (7 c+74)\beta^2]}{1360800 \beta^{10} L^2}. \nn
\eea
\end{widetext}
It is easy to see that
\be
\sum_i p_i \Xrhoi = \lag \cX \rag_\b, ~~ \mX=T,\cA.
\ee
We have also used
\be \label{SipiTirXi}
\sum_i p_i \Trhoi^r X_i = \Big( \f{2\pi}{L} \Big)^r
                          \f{\p_\b^r(\ep^{\f{\pi c L}{12\b}} \sum_i p_i X_i )}
                            {\ep^{\f{\pi c L}{12\b}}},
\ee
where $r$ is an arbitral integer and $X_i$ can be either the one-point function of an operator or a product of the one-point functions. We also have
\be \label{SipiXiYi}
\sum_i p_i \Xrhoi \Yrhoi = \f{1}{L} \!\int_{-L/2}^{L/2}\!\! d x \lag \mX(x)\mY(0) \rag_\b,
\ee
with $\mX=T,\cA$, $\mY=T,\cA$. This follows from the fact that both $T$ and $\mA$ are KdV currents that commute with each other and we can choose the states $|i\rag$ as the common eigenstates of their zero modes. Explicitly, we derive (\ref{SipiXiYi}) as follows. On a torus $\mT$ with spatial period $L$ and temporal period $\b$ there is the two-point function
\begin{widetext}
\be
\lag \mX(x) \mY(0) \rag_\mT =
\f{1}{Z(\b)} \sum_{i,i'} \ep^{-\f{2\pi \b}{L}( \D_i - \f{c}{12} )} \ep^{\f{2\pi \ii x}{L}(s_{i'}-s_i)}
\lag i | \mX | i' \rag \lag i' | \mY | i \rag.
\ee
For bosonic $\mX$, $\mY$, we require that $s_{i'}-s_i$ is an integer for $\lag i | \mX | i' \rag \lag i' | \mY | i \rag$ being non-vanishing. Then we get
\be
\f{1}{L} \!\int_{-L/2}^{L/2}\!  d x \lag \mX(x) \mY(0) \rag_\mT =
\f{1}{Z(\b)} \sum_{i,i'} \ep^{-\f{2\pi \b}{L}( \D_i - \f{c}{12} )} \d_{s_{i'} s_i}
\lag i | \mX | i' \rag \lag i' | \mY | i \rag.
\ee
\end{widetext}
For $\mX$, $\cY$ being operators in the vacuum conformal family, we require that $|i\rag$ and $|i'\rag$ are in the same conformal family.
The delta function $\d_{s_{i'} s_i}$ further requires that $|i\rag$ and $|i'\rag$ are at the same level, and so only the zero modes of $\mX$, $\cY$ contribute to $\lag i | \mX | i' \rag \lag i' | \mY | i \rag$.
For $\mX$, $\mY$ being KdV currents, the states $|i\rag$ can be organized as the common eigenstates of their zero modes. Then we have $|i\rag = | i' \rag$.
Omitting the exponentially suppressed terms in high temperature limit, we have $\lag \mX(x) \mX(0) \rag_\mT =\lag \mX(x) \mX(0) \rag_\b$.
We finally arrive at (\ref{SipiXiYi}). By omitting the exponentially suppressed terms and by an analytical continuation, in evaluating  (\ref{SipiXiYi}) we use the integral
\be
\f{1}{L} \!\int_{-L/2}^{L/2}\! \f{dx}{( \sinh\f{\pi x}{\b} )^S} = \f{\b}{L} \f{\G(\f{S}{2})\G(\f{1-S}{2})}{\pi^{\f32}}.
\ee
As consistency checks, we get the same $\sum_i p_i T_i^2$, $\sum_i p_i T_i \mA_i$ from (\ref{SipiTirXi}) and (\ref{SipiXiYi}).

Using (\ref{SAOPE}) and (\ref{SipiXi}), we get the average EE, (13) in the main text,
\begin{widetext}
\bea \label{SipiSice}
&& \sum_i p_i S_{A,i} = \f{c}{3}\log\f{\ell}{\e}
  +\frac{\pi^2 c \ell^2}{18 \beta^2}
  -\frac{\pi^3\ell^4(\pi  c L +24 \beta)}{540 \beta^4 L}
  +\frac{\pi^4\ell^6 (\pi^2 c^2 L^2+72 \pi c \beta L+864 \beta^2)}{8505 c \beta^6 L^2} \nn\\
&& \phantom{\sum_i p_i S_{A,i} =}
  +\frac{\pi^5\ell^8 [-7\pi^3 c^3 L^3-2160\pi^2 c^2\beta L^2+1120\pi c(c-28)\beta^2 L-80640(c+8)\beta^3]}{793800c^2\beta^8L^3} \nn\\
&& \phantom{\sum_i p_i S_{A,i} =}
  +\frac{\pi^6 \ell^{10}}{9823275 c^3 \beta^{10} L^4}
   [7 \pi^4 c^4 L^4+8592 \pi^3 c^3 \beta  L^3-6720 \pi^2 c^2(c-100) \beta^2 L^2\nn\\
&& \phantom{\sum_i p_i S_{A,i} =}
    +2903040 \pi c \beta^3 L+29030400(c+5) \beta^4]
   +O(\ell^{12}).
\eea
and then the short interval Holevo information, (14) in the main text,
\bea
&& \chi_A = \frac{2 \pi^3 \ell^4}{45 \beta^3 L}
               -\frac{8\pi^4\ell^6 (\pi  c L + 12 \beta )}{945 c \beta^5 L^2}
               +\frac{ 2 \pi^5\ell^8[27 \pi^2 c^2 L^2-14 \pi c(c-28) \beta L+1008(c+8) \beta^2 ]}{19845 c^2 \beta^7 L^3}\\
&& \phantom{\chi_A =}
               +\frac{16 \pi^6 \ell^{10}[ -179 \pi^3 c^3 L^3+140 \pi^2 c^2 (c-100) \beta L^2-60480 \pi c  \beta^2 L -604800(c+5) \beta^3 ] }{3274425 c^3 \beta^9 L^4}
               +O(\ell^{12}). \nn
\eea
At order $\ell^{12}$, the quasiprimary operators at level six that are not currents of the KdV charges begin to contribute, and the above calculation method breaks down. Now we do not know how to solve this technical problem.

The mutual information of two disjoint intervals with cross ratio $x$ on a complex plane can be organized by orders of large $c$ as
\be
I(x) = I_\rL(x) + I_\NL(x) + \cdots.
\ee
The leading part of the mutual information is universal \cite{Headrick:2010zt}
\be
I_\rL(x) = \lt\{ \ba{ll}
0 & x<1/2 \\
\df{c}{3} \log\df{x}{1-x} & x>1/2
\ea\rt.\!\!\!.
\ee
The remaining part of the mutual information satisfies $I_\NL(x)=I_\NL(1-x)$, $\cdots$. With contributions of only the vacuum conformal family, we have \cite{Barrella:2013wja,Chen:2013dxa,Beccaria:2014lqa,Li:2016pwu}
\be
I_\NL(x) = \frac{x^4}{630}+\frac{2 x^5}{693}+\frac{15 x^6}{4004}+\frac{x^7}{234}+\frac{167 x^8}{36036}+\frac{69422 x^9}{14549535}+\frac{122x^{10}}{24871}+O(x^{11}).
\ee
In principle, one can use the methods in \cite{Barrella:2013wja,Chen:2013dxa,Beccaria:2014lqa,Li:2016pwu} and calculate this mutual information to higher orders of $\ell$, but in practice it would be very involved and has not been done yet.

Finally, we get the long interval Holevo information, (16) in the main text,
\bea
&& \chi_B = \frac{\pi  c L}{3 \beta }
          -\frac{2 \pi^3\ell^4 ( 4 \pi  L - 7 \beta )}{315 \beta^4 L}
          +\frac{32 \pi^5 \ell^5}{3465 \beta^5}
          +\frac{8 \pi^4\ell^6 ( 32 \pi^2 L^2-143 \pi \beta L )}{135135 \beta^6L^2}
          -\frac{32 \pi^7 \ell^7}{27027 \beta^7}\nn\\
&& \phantom{\chi_B =}
          +\frac{2 \pi^5\ell^8 ( -109116 \pi^3 L^3+19305 \pi^2 \beta L^2
                                 -10010 \pi \beta^2 L )}{14189175 \beta^8 L^3}
          +\frac{2596928 \pi^9 \ell^9}{26189163 \beta^9} \nn\\
&& \phantom{\chi_B =}
          +\frac{16 \pi^6\ell^{10} ( -369060974 \pi^4 L^4
             -751621 \pi^3 \beta  L^3
             +587860 \pi^2 \beta^2 L^2)}{13749310575 \beta^{10} L^4}
          +O(\ell^{11},1/c).
\eea
\end{widetext}

\begin{center}
{\normalsize \textbf{Canonical ensemble thermal state with low temperature}}
\end{center}

For the canonical ensemble thermal state with low temperature, we only consider the contributions from the holomorphic sector of the vacuum conformal family. The CFT is on a torus $\mT$ with spatial period $L$ and temporal period $\b$. In low temptation limit $L\ll\b$, and to get non-vanishing corrections to the Holevo information we have to include the exponentially suppressed terms by $q=\ep^{-\f{2\pi\b}{L}} \ll 1$.

The holomorphic part of the partition function is
\be
Z(q) = q^{-\f{c}{24}} [ 1+q^2+q^3+2 q^4+O(q^5) ].
\ee
Similar to (\ref{SipiTirXi}), the average products of one-point functions for the stress tensor $T$ can be written as
\be
\sum_i p_i \lag T \rag_{\r_i}^r = \Big( \f{2\pi\ii}{L} \Big)^{2r} \f{(q\p_q)^r Z(q)}{Z(q)}.
\ee
We get the results
 \begin{widetext}
\bea
&& \sum_i p_i \lag T \rag_{\r_i} = \lag T \rag_\cT
                                 = \frac{\pi ^2 c}{6 L^2}
                                  -\frac{8 \pi ^2 q^2}{L^2}
                                  -\frac{12 \pi ^2 q^3}{L^2}
                                  -\frac{24 \pi ^2 q^4}{L^2}
                                  +O(q^5) \nn\\
&& \sum_i p_i \lag T \rag_{\r_i}^2 = \frac{\pi ^4 c^2}{36 L^4}
                                    -\frac{8 \pi ^4 (c-24) q^2}{3 L^4}
                                    -\frac{4 \pi ^4 (c-36) q^3}{L^4}
                                    -\frac{8\pi ^4 (c-56) q^4}{L^4}
                                    +O(q^5)  \\
&& \sum_i p_i \lag T \rag_{\r_i}^3 = \frac{\pi ^6 c^3}{216 L^6}
                                    -\frac{2 \pi ^6 (c^c-48c+768) q^2}{3 L^6}
                                    -\frac{\pi ^6 ( c^2-72c +1728) q^3}{L^6}
                                    -\frac{2 \pi ^6 (c^2-112c +3840) q^4}{L^6}
                                    +O(q^5), \nn
\eea
from which we get the short interval EE
\bea
&& S_A = \f {c} {6}\log\f {\ell} {\e}
+ \Big( -\frac {c} {36}
        + \frac {4 q^2} {3}
        + 2 q^3
        + 4 q^4
        + O(q^5) \Big) \Big( \f {\pi\ell} {L} \Big)^2
+ \Big( - \frac {c} {1080}
        + \frac {4 q^2} {45}
        + \frac {2 q^3} {15}
        + \frac {4 (c - 8) q^4} {15 c}\nn\\
&& \phantom{S_A =}
        + O(q^5) \Big) \Big( \f {\pi\ell} {L} \Big)^4
+ \Big( - \frac {c} {17010}
        + \frac {8 q^2} {945}
        + \frac {4 q^3}{315}
        + \frac {8 (c - 16) q^4} {315 c}
        + O(q^5) \Big) \Big( \f {\pi\ell} {L} \Big)^6
+ O (\ell^8),
\eea
and average EE
\bea
&& \sum_i p_i S_{A,i} = \f {c} {6}\log\f {\ell} {\e}
+ \Big ( - \frac {c} {36}
         + \frac {4 q^2} {3}
         + 2 q^3
         + 4 q^4
         + O (q^5) \Big) \Big ( \f {\pi\ell} {L} \Big)^2
+ \Big (  -\frac {c} {1080}
         + \frac {4 (c - 24) q^2} {45 c} \nn\\
&& \phantom{\sum_i p_i S_{A,i} =}
         + \frac {2 (c - 36) q^3} {15 c}
         + \frac {4 (c - 56) q^4} {15 c}
         + O (q^5) \Big) \Big ( \f {\pi\ell} {L} \Big)^4
+ \Big ( - \frac {c} {17010}
         + \frac {8 (c^c-48c+768) q^2} {945 c^2} \nn\\
&& \phantom{\sum_i p_i S_{A,i} =}
         + \frac {4 ( c^2-72c +1728) q^3} {315 c^2}
         + \frac {8 (c^2-112c +3840) q^4} {315 c^2}
         + O (q^5) \Big) \Big ( \f {\pi\ell} {L} \Big)^6
         + O (\ell^8).
\eea
Then we get the short interval Holevo information (21) in the main text. %(\ref{chiAlow}).

The low temperature long interval EE has been calculated in \cite{Chen:2017ahf}
\bea
&& S_B = \Big[ \Big( \f{4\pi\b}{L} +1 \Big) q^2 + O(q^3) \Big]
+ \f {c} {6}\log\f {\ell} {\e}
+ \Big[ -\frac {c} {36}
        + \frac {4 q^2} {3}
        + O(q^3) \Big] \Big( \f {\pi\ell} {L} \Big)^2 \nn\\
&& \phantom{S_B=}
+ \Big[ - \frac {c} {1080}
        + \frac {4(c-24)q^2} {45c}
        - \f{32\pi\b(\b^2+L^2)(4\b^2+L^2)}{15L^5}q^2
        + O(q^3) \Big] \Big( \f {\pi\ell} {L} \Big)^4 + O(\ell^5).
\eea
\end{widetext}
Noting the thermal entropy
\be
S(L) = \Big( \f{4\pi\b}{L} +1 \Big) q^2 + O(q^3),
\ee
we get the long interval Holevo information (22) in the main text. %(\ref{chiBlow}).

\begin{center}
{\normalsize \textbf{Microcanonical ensemble thermal state}}
\end{center}

The density of states at fixed energy $E$ is defined as
\be
\O(E) = \sum_i \d(E-E_i).
\ee
The energy $E$ can be written in terms of the scaling dimension as $E = \f{2\pi}{L} ( \D - \f{c}{12} )$. For the ground state $\D=0$, and so $E = -\f{\pi c}{6L}$. In a unitary CFT $\D \geq 0$, and so $E \geq -\f{\pi c}{6L}$.
The canonical ensemble partition function can be written as
\be
Z(\b) = \sum_i \ep^{-\b E_i} = \int_{-\f{\pi c}{6L}}^{+\inf} d E \ep^{-\b E} \O(E).
\ee
Then one can use the inverse Laplace transformation to get the density of states
\be \label{OEint}
\O(E) = \f{1}{2\pi\ii} \int_{\g-\ii\inf}^{\g+\ii\inf} d \b \ep^{\b E} Z(\b).
\ee
We omit the exponentially suppressed term by higher energy, or equivalently high temperature, and have $Z(\b) = \ep^{\f{\pi c L}{6\b}}$. Beyond the saddle point approximation in  \cite{Cardy:1986ie,Carlip:2000nv}, the integral (\ref{OEint}) leads to
\be
\O(E) = \sqrt{\f{\pi c L}{6E}} I_1\Big(\sqrt{\f{2\pi c L E}{3}}\Big).
\ee

As what have been done in \cite{Kraus:2016nwo,Brehm:2018ipf,Romero-Bermudez:2018dim,Hikida:2018khg}, for other general canonical ensemble average in the form
\be
\mO(\b) = \f{1}{Z(\b)} \sum_i \ep^{-\b E_i} \mO_i,
\ee
we can also do an inverse Laplace transformation and get the microcanonical ensemble average
\be
\mO(E) = \f{1}{\O(E)} \f{1}{2\pi\ii}
          \int_{\g-\ii\inf}^{\g+\ii\inf}
          d \b \ep^{\b E} \mO(\b)Z(\b).
\ee
Note that $\mO_i$ can be any quantity defined for the pure state $|i\rag$, e.g., a one-point function, a product of one-point functions, and the EE.
For the case that the canonical average $\mO(\b)$ is a polynomial of $\b$, we get the microcanonical average $\mO(E)$ from $\mO(\b)$ by the substitute
\be \label{substitute1}
\b^{-k} \to \Big( \f{\pi c L}{6E} \Big)^{-k/2}
            \f{I_{k-1} (\sqrt{\f{2\pi c L E}{3}})}
              {I_1 (\sqrt{\f{2\pi c L E}{3}})}.
\ee
It is convenient to define the effective length scale
\be
\l := \sqrt{\f{\pi c L}{6E}},
\ee
and the substitute (\ref{substitute1}) becomes
\be \label{substitute}
\b^{-k} \to \l^{-k}
            \f{I_{k-1} ( \f{\pi c L}{3\l} )}
              {I_1 ( \f{\pi c L}{3\l} )}.
\ee
In the following, we just use the shorthand notation $I_\n$ for  $I_\n( \f{\pi c L}{3\l} )$.

Using the substitute (\ref{substitute}), we can get the microcanonical ensemble one-point functions from the canonical ensemble ones
\be
\lag T \rag_E = -\frac{\pi ^2 c}{6 \lambda ^2}, ~~
\lag \mA \rag_E = \frac{\pi ^4 c (5 c+22) {I_3}}{180 \lambda ^4 {I_1}},
\ee
and then we get the short interval EE
\begin{widetext}
\bea \label{SAE}
&& S_A = \f{c}{3} \log \f{\ell}{\e}
       + \frac{\pi ^2 c \ell ^2}{18 \lambda ^2}
       - \frac{ \pi ^4 c \ell ^4}{540 \lambda ^4}
       + \frac{\pi ^6 c \ell ^6}{8505 \lambda^6}
       - \frac{\pi ^8 c \ell ^8 [ 5 (c+8) {I_1^2}-2 (5 c+22) {I_1} {I_3}+(5 c+22) {I_3^2} ] }
              {2041200 \lambda ^8 {I_1^2}} \nn\\
&& \phantom{S_A =}
       + \frac{\pi ^{10} c \ell ^{10} [ 5 (c+5) {I_1^2}-2 (5 c+22) {I_1} {I_3}+(5 c+22){I_3^2}]}
             {4209975 \lambda ^{10}{I_1^2}}
       + O(\ell ^{12}).
\eea
Note that the result is valid with the exponentially suppressed terms of high energy being omitted and the power suppressed terms being kept. From the average EE in high temperature canonical ensemble thermal state (\ref{SipiSice}), we get the average EE in high energy microcanonical ensemble thermal state
\bea
&& \sum_i p_i S_{A,i} = \f{c}{3}\log\f{\ell}{\e}
  +\frac{\pi^2 c \ell^2}{18 \l^2}
  -\frac{\pi^3\ell^4(\pi  c L I_3 +24 \l I_2)}{540 \l^4 L I_1}
  +\frac{\pi^4 \ell^6(\pi^2 c^2 L^2 I_5 +72 \pi c \l L I_4 +864 \l^2 I_3)}
        {8505 c \l^6 L^2 I_1} \nn\\
&& \phantom{\sum_i p_i S_{A,i} =}
  +\frac{\pi^5\ell^8 [-7\pi^3 c^3 L^3 I_7 -2160\pi^2 c^2\l L^2 I_6 +1120\pi c(c-28)\l^2 L I_5 -80640(c+8)\l^3 I_4 ]}
        {793800c^2\l^8L^3 I_1} \nn\\
&& \phantom{\sum_i p_i S_{A,i} =}
  +\frac{\pi^6 \ell^{10}}{9823275 c^3 \l^{10} L^4 I_1}
   [7 \pi^4 c^4 L^4 I_9 +8592 \pi^3 c^3 \l  L^3 I_8 -6720 \pi^2 c^2(c-100) \l^2 L^2 I_7 \nn\\
&& \phantom{\sum_i p_i S_{A,i} =}
    +2903040 \pi c \l^3 L I_6 +29030400(c+5) \l^4 I_5]
   +O(\ell^{12}).
\eea
Then we get the short interval Holevo information, (25) in the main text,
\bea
&& \chi_A =
 \frac{\pi ^3 \ell ^4[ \pi  c L ({I_3}-{I_1})+24 \lambda{I_2} ]}
      {540\lambda ^4 L{I_1}}
-\frac{\pi ^4 \ell ^6 [ \pi ^2 c^2 L^2 ({I_5}-{I_1})+72 \pi c \lambda  L {I_4} +864 \lambda^2 {I_3} ]}
      {8505 c\lambda ^6 L^2{I_1}} \nn\\
&& \phantom{\chi_A =}
+\frac{\pi ^5 \ell ^8}
      {14288400 c^2 {I_1}^2\lambda ^8 L^3}
 \{  -7 \pi ^3 c^3 L^3 [ 5 (c+8) {I_1^2} - (5 c+22) {I_3} (2 {I_1} - {I_3})-18 {I_1} {I_7} ]
     +38880\pi ^2 c^2 \lambda  L^2{I_1} {I_6} \nn\\
&& \phantom{\chi_A =}
     +1451520 (c+8) \lambda^3{I_1} {I_4}
     -20160 \pi  (c-28) c \lambda ^2 L{I_1} {I_5} \} \nn\\
&& \phantom{\chi_A =}
+\frac{\pi ^6 \ell ^{10}}{29469825 c^3\lambda ^{10} L^4{I_1^2}}
 \{ 7 \pi ^4 c^4 L^4 [ 5 (c+5){I_1^2}-(5 c+22){I_3} (2{I_1}-{I_3})-3{I_1}{I_9} ]
   -25776\pi ^3 c^3 \lambda  L^3{I_1} {I_8} \nn\\
&& \phantom{\chi_A =}
   +20160 \pi ^2 (c-100) c^2\lambda ^2 L^2{I_1} {I_7}
   -87091200 (c+5) \lambda ^4{I_1} {I_5}
   -8709120 \pi  c \lambda ^3 L {I_1}{I_6} \}
+O(\ell^{12}).
\eea
\end{widetext}

As a byproduct in the letter, we can show that the reduced density matrix of the high energy microcanonical ensemble thermal state $\r_{A,E}$ equals the reduced density matrix of the high temperature canonical ensemble thermal state $\r_{A,\b}$ in the thermodynamic limit, or equivalently high temperature limit. The difference of the two reduced density matrices are power suppressed. We stress that this result does not depend on the large $c$ limit and applies to any 2D CFT.

In the first step, we identify the energy expectation values of the two states, and so we have $\l =\b$. To make it more concrete, in the following we will show that the EEs of the two states are the same up to power corrections
\be \label{SAEmSAb}
S_{A,E} - S_{A,\b} = O\Big(\f{\l}{L},\f{\ell}{L}\Big),
\ee
and the relative entropies of the two reduced density matrixes are also power suppressed
\bea \label{SrAErAb}
&& S(\r_{A,E} \| \r_{A,\b}) = O\Big(\f{\l}{L},\f{\ell}{L}\Big), \nn\\
&& S(\r_{A,\b} \| \r_{A,E}) = O\Big(\f{\l}{L},\f{\ell}{L}\Big).
\eea

Using the modular transformation of one-point functions on a torus, one can calculate the average one-point function of a general quasiprimary operator $\mX$ with scaling dimension $\D_\mX$ and spin $s_\mX$ in the microcanonical ensemble thermal state as \cite{Kraus:2016nwo}
\bea \label{XE}
&& \lag \mX \rag_E \approx \f{\lag \mY|\mX|\mY\rag}{\ii^{s_\mX}}
                        \Big( \f{6L}{\pi c} \Big)^{\f14}
                        ( -E_\mY )^{-\f{\D_\mX}{2}+\f14}
                        E^{\f{\D_\mX}{2}} \nn\\
&& \phantom{\lag \mX \rag_E \approx} \times
                        \ep^{- ( \sqrt{\f{2\pi c}{3}} - 2\sqrt{-L E_\mY}  )\sqrt{L E}},
\eea
with $\mY$ being a quasiprimary operator with the lowest scaling dimension that satisfies $\lag \mY|\mX|\mY\rag \neq 0$. Note that $E_\mY \geq -\f{\pi c}{6L}$, and it is assumed that $\mY$ is not so heavy so that $E_\mY < 0$. In the derivation of (\ref{XE}) the saddle point approximation has been used and the power suppressed terms by large $E$ has been omitted. As a consistency check of the normalization of (\ref{XE}), we can see that for the identity operator $\mX=1$, we have $\mY=1$, and the right hand side of (\ref{XE}) is 1.

When $\mX$ is in a non-vacuum conformal family, we have $E_\mY > -\f{\pi c}{6L}$ and the one point function $\lag \mX \rag_E$ is exponentially suppressed, and thus can be omitted. When $\mX$ is in the vacuum conformal family, $\mY$ is the identity operator, or in other words the state $|\mY\rag = \mY(0)|0\rag$ is just the ground sate $|0\rag$. For this case, without loss of generality we choose $\mX$ to be holomorphic, and so $\D_\mX=s_\mX=h_\mX$ is an integer. Noting that $\lag 0 |\mX| 0 \rag = \lag \mX\rag_L$, $E_0 = -\f{\pi c}{6L}$, $E=\f{\pi c L}{6\l^2}$, we use (\ref{XE}) and get
\be
\lag \mX \rag_E \approx \lag \mX \rag_L \Big( \f{L}{\ii\l} \Big)^{h_\mX}.
\ee

For the high temperature canonical ensemble thermal state, we also omit the exponentially suppressed terms. When $\mX$ is in a non-vacuum conformal family we have $\lag \mX \rag_\b=0$. When $\mX$ is in the holomorphic vacuum conformal family, we have
\be
\lag \mX \rag_\b = \lag \mX \rag_L \Big( \f{L}{\ii\b} \Big)^{h_\mX}.
\ee
Since we have identified $\l=\b$, we get that for all quasiprimary operators
\be \label{XEXb}
\lag \mX \rag_E = \lag \mX \rag_\b + O\Big(\f{\l}{L}\Big)\;.
\ee
The equivalence (\ref{XEXb}) is exact under the thermodynamic limit in \cite{Lashkari:2016vgj,Dymarsky:2016aqv}, i.e., $E \to \inf$ and $L \to \inf$ with $E/L$ being finite. The equality of the two reduced density matrices are expected to be valid for general $\ell$, $\l=\b$ as long as the thermodynamic limit is taken $\ell/L \to 0$, $\l/L \to 0$.

Using OPE of twist operators, one can write the EE and relative entropy as sums of products of one-point functions, and the coefficients of the products are universal and do not depend on parameters of the state. Then we use (\ref{XEXb}) and get the relations (\ref{SAEmSAb}), (\ref{SrAErAb}).

\begin{figure}[htpb]
  \centering
  \includegraphics[height=0.22\textwidth]{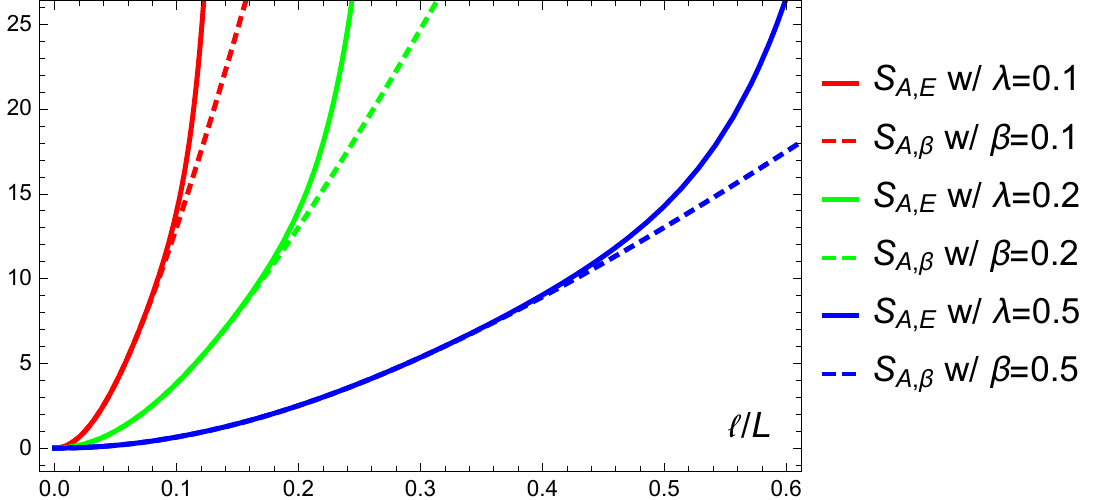}
  \caption{EEs of the high energy microcanonical ensemble thermal state and the high energy canonical ensemble thermal state. We have omitted the divergent part $\f{c}{3}\log\f{\ell}{\e}$ and set $c=30$.}\label{com}
\end{figure}

\begin{figure}[htpb]
  \centering
  \includegraphics[height=0.22\textwidth]{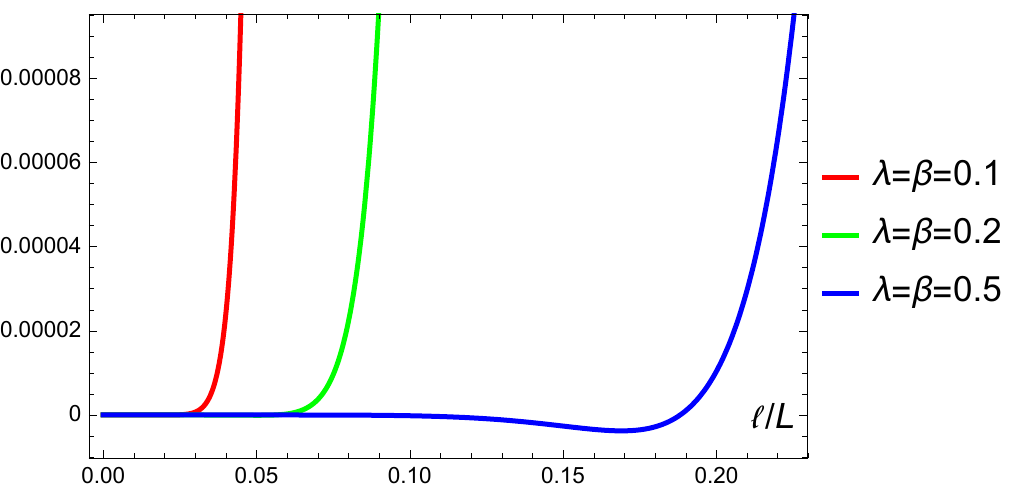}
  \caption{The EE difference of the microcanonical ensemble and canonical ensemble thermal states $S_{A,E}-S_{A,\b}$. We have set $c=30$.}\label{deltaS}
\end{figure}

It is interesting to compare directly the high energy microcanonical ensemble thermal state EE $S_{A,E}$ (\ref{SAE}),
in which the exponentially suppressed terms are omitted and but the power suppressed terms are kept,
and the high temperature canonical ensemble thermal state EE $S_{A,\b} = \f{c}{3}\log ( \f{\b}{\pi\e} \sinh\f{\pi\ell}{\b} )$, in which the exponentially suppressed terms are omitted and there are no power suppressed terms.
We plot them in Fig.~\ref{com}, and their difference in Fig.~\ref{deltaS}.
We can see the EEs of the two states are very close as long as $\ell<\b$, and the large difference at $\ell>\b$ can be attributed to the breaking down of the short interval expansion in (\ref{SAE}).

To get the long interval Holevo information, we need to calculate the long interval EE in the microcanonical ensemble thermal state.
The relevant states $|i\rag$ are at the same energy and are the common eigenstates of the zero modes of $T$ and $\mA$, but they are not necessarily the eigenstates of the zero modes of level six quasiprimary operators $\mB$ and $\mD$, whose definitions can be found in \cite{Chen:2013kpa,Chen:2013dxa}. We use the OPE of twist operators for a long interval \cite{Chen:2014ehg,Chen:2015kua,Chen:2017ahf} and get the partition function
\blea \label{trBrBn}
&& \tr_B \r_B^n = \Big( \f{\ell}{\e} \Big)^{-4h_\s} \f{1}{\O^{n-1}}
\Big\{ 1 +
\f{2}{\O}
\sum_i{}' \big[
         \ell^2 b_T \Trhoi
       + \ell^4 b_{TT} \Trhoi^2
       + \ell^6 b_{TTT} \Trhoi^3 \nn\\
&& \phantom{\tr_B \r_B^n =}
       + \ell^8 \big( b_{\mA\mA} \Arhoi^2
                    + b_{TT\mA} \Trhoi^2\Arhoi
                    + b_{TTTT} \Trhoi^4 \big) \nn\\
&& \phantom{\tr_B \r_B^n =}
       + \ell^{10} \big( b_{T\mA\mA} \Trhoi\Arhoi^2
                       + b_{TTT\mA} \Trhoi^3\Arhoi
                       + b_{TTTTT} \Trhoi^5 \big) \nn\\
&& \phantom{\tr_B \r_B^n =}
       + \ell^{12} \big( b_{\mB\mB} \Brhoi^2
                       + b_{\mD\mD} \Drhoi^2
                       + b_{T\mA\mB} \Trhoi\Arhoi\Brhoi
                       + b_{T\mA\mD} \Trhoi\Arhoi\Drhoi \nn\\
&& \phantom{\tr_B \r_B^n =}
                       + b_{\mA\mA\mA} \Arhoi^3
                       + b_{TTT\mB} \Trhoi^3\Brhoi
                       + b_{TTT\mD} \Trhoi^3\Drhoi
                       + b_{TT\mA\mA} \Trhoi^2\Arhoi^2 \nn\\
&& \phantom{\tr_B \r_B^n =}
                       + b_{TTTT\mA} \Trhoi^4\Arhoi
                       + b_{TTTTTT} \Trhoi^6  \big) \big] \nn\\
&& \phantom{\tr_B \r_B^n =}
   + \f{2\ell^{12}}{\O}
     \sum_{i\neq i'}{}'\big(  b_{\mB\mB} \lag i | \mB | i' \rag \lag i' | \mB | i \rag
                         + b_{\mD\mD} \lag i | \mD | i' \rag \lag i' | \mD | i \rag \big)
+O[\ell^{14},(n-1)^2] \Big\}.
\elea
The conformal weight of the twist operators is $h_\s = \f{c(n^2-1)}{24n}$ \cite{Calabrese:2004eu}.
There are contributions from both the holomorphic and anti-holomorphic sectors of the vacuum conformal family, and we only consider the case that the contributions from the holomorphic and anti-holomorphic sectors are exactly the same.
We have restricted the sum $\sum{}'$ to the states of the fixed energy $E$, and $\O$ is the total number of such states $\O = \sum_i'1$.
The coefficients $b_{\mX_1 \cdots \mX_k}$ are defined from the OPE coefficients $d_{\mX_1 \cdots \mX_k}^{j_1 \cdots j_k}$ of the quasiprimary operators $\mX_1^{j_1} \cdots \mX_k^{j_k}$ in the $n$-fold CFT \cite{Chen:2016lbu}, and their explicit forms are not important to us.
We have also used the results in \cite{Guo:2018pvi} and omitted some order $(n-1)^2$ terms in (\ref{trBrBn}) in the $n \to 1$ limit. The omitted terms are relevant to the R\'enyi entropy but are irrelevant to the EE.
Then the long interval EE can be written as
\bea
&& S_B = \log \O
+ \f{1}{\O} \sum_i{}' S_{A,i}
   + \f{2\ell^{12}}{\O}
     \sum_{i\neq i'}{}'\big(  a_{\mB\mB} \lag i | \mB | i' \rag \lag i' | \mB | i \rag \nn\\
&& \phantom{S_B =}
                         + a_{\mD\mD} \lag i | \mD | i' \rag \lag i' | \mD | i \rag \big)
+O(\ell^{14}),
\eea
with the coefficients \cite{Guo:2018pvi}
\bea
&& a_{\mB\mB} = -\frac{25}{123552 c (70 c+29)}, \\
&& a_{\mD\mD} = -\frac{70 c+29}{18018 c (2 c-1) (5 c+22) (7 c+68)}. \nn
\eea
We note that the thermal entropy is $S(L)=\log\O$ and get the long interval Holevo information, (26) in the main text
\be
\chi_B - S(L) = O(\ell^{12}).
\ee
Note that neither $\mB$ nor $\mD$ is a current of the KdV charges.
We do not know how to calculate $\sum_{i\neq i'}' \lag i | \mB | i' \rag \lag i' | \mB | i \rag$ or $\sum_{i\neq i'}' \lag i | \mD | i' \rag \lag i' | \mD | i \rag$, and so we cannot evaluate the order $\ell^{12}$ part of the long interval Holevo information.

\begin{center}
{\normalsize \textbf{Contributions from a non-identity primary operator}}
\end{center}

Similar to what we have done for the contributions to the Holevo information from the vacuum conformal family, we can use the OPE of twist operators \cite{Calabrese:2004eu,Cardy:2007mb,Headrick:2010zt,Calabrese:2010he,Rajabpour:2011pt,Chen:2013kpa,%
Chen:2014ehg,Bianchini:2015uea,Chen:2015kua,Chen:2017ahf}, and get the leading contributions from a non-identity primary operator $\psi$, (27) in the main text.

For the canonical ensemble thermal state in the high temperature limit, we can further write the results with the exponentially suppressed terms omitted as
\begin{widetext}
\bea
&& \d_\psi \chi_A = \f{\sr\pi\G(\D_\psi+1)\ell^{2\D_\psi}}{2^{2\D_\psi+2}\G(\D_\psi+\f32)}
                 \f{\ii^{2s_\psi}}{\a_\psi}
                 \f{1}{Z(\b)} \sum_i \ep^{-\b E_i} \lag \psi \rag_{\r_i}^2
                 + o(\ell^{2\D_\psi}), \nn\\
&& \d_\psi \chi_B =
               - \f{\ell^{2\D_\psi}}{2^{2\D_\psi+1}}
                 \f{\ii^{2s_\psi}}{\a_\psi}
                 \f{1}{Z(\b)} \sum_{i \neq i'}
                 \lag i | \psi | i' \rag \lag i' | \psi | i \rag
                 \ep^{-\b E_i}
                 \p_n \Big[ \sum_{j=1}^{n-1} \f{\ep^{j \b ( E_i - E_{i'} )}}{(\sin \f{\pi j}{n})^{2\D_\psi}} \Big]_{n=1}
                 + o(\ell^{2\D_\psi}).
\eea

For the canonical ensemble thermal state in the low temperature limit, we get
\bea
&& \d_{\psi,\psi'} \chi_A = \f{\sr\pi\G(\D_\psi+1)\ell^{2\D_\psi}q^{\D_{\psi'}}}{2^{2\D_\psi+2}\G(\D_\psi+\f32)}
                 \f{\ii^{2s_\psi}}{\a_\psi\a_{\psi'}^2}
                 \lag \psi' | \psi | \psi' \rag^2
                 + o(\ell^{2\D_\psi},q^{\D_{\psi'}}), \nn\\
&& \d_\psi \chi_B = \d_\psi S(L)
                  - \Big( \f{\pi\ell}{L} \Big)^{2\D_\psi} \ii^{4s_\psi}
                    \p_n \Big[ \sum_{j=1}^{n-1} \f{q^{j \D_\psi}}{(\sin \f{\pi j}{n})^{2\D_\psi}} \Big]_{n=1}
                  + o(\ell^{2\D_\psi},q^{\D_{\psi}}).
\eea
Here $\psi'$ is the primary operator with least conformal dimension that satisfies $\lag \psi' | \psi | \psi' \rag \neq 0$. Note also that
\be
\d_\psi S(L) = \Big( \f{2\pi\D_\psi \b}{L} + 1 \Big) q^{\D_\psi} + o(q^{\D_\psi}).
\ee

For the microcanonical ensemble thermal state, we get
\bea
&& \d_\psi \chi_A = \f{\sr\pi\G(\D_\psi+1)\ell^{2\D_\psi}}{2^{2\D_\psi+2}\G(\D_\psi+\f32)}
                    \f{\ii^{2s_\psi}}{\a_\psi}
                    \Big[ \f{1}{\O}\sum_i{}' \lag \psi \rag_{\r_i}^2
                        - \f{1}{\O^2}\Big( \sum_i{}'\lag \psi \rag_{\r_i} \Big)^2
                    \Big]
                    + o(\ell^{2\D_\psi}), \nn\\
&& \d_\psi \chi_B = - \f{\sr\pi\G(\D_\psi+1)\ell^{2\D_\psi}}{2^{2\D_\psi+2}\G(\D_\psi+\f32)}
                      \f{\ii^{2s_\psi}}{\a_\psi}
                      \f{1}{\O} \sum_{i\neq i'}{}' \lag i | \psi | i' \rag \lag i' | \psi | i \rag
                    + o(\ell^{2\D_\psi}).
\eea
These results are not universal, and we cannot evaluate them without knowing details of the CFT.
\end{widetext}

%%%%%%%%%%%%%%%%%%%%%%%%%%%%%%%%%%%%%%%%%%%%%%%%%%%%%%%%%%%%%%%%%%
%% End supplemental material
%%%%%%%%%%%%%%%%%%%%%%%%%%%%%%%%%%%%%%%%%%%%%%%%%%%%%%%%%%%%%%%%%%

\end{document}